\documentclass[preprintnumbers,amsmath,amssymb,twocolumn]{revtex4}
\usepackage[colorlinks=true,citecolor=red,filecolor=green,linkcolor=blue,pdfnewwindow=true]{hyperref}
\usepackage{graphicx,ulem}
\usepackage{makeidx}
\usepackage{mathrsfs}
\usepackage{threeparttable}
\newcommand{\stkout}[1]{\ifmmode\text{\sout{\ensuremath{#1}}}\else\sout{#1}\fi}
\usepackage{empheq}
\usepackage{csquotes}
\usepackage{soul}
\newcommand{\di}{\mathrm{i}}
\usepackage[caption=false]{subfig}
\usepackage[title]{appendix}

\def\prl#1#2#3{{Phys. Rev. Lett.} {\bf #1}, #2 (#3)} 

\def\ijbc#1#2#3{Int J. Bifurcation and Chaos {\bf #1}, #2 (#3)}
\def\fop#1#2#3{Front. Phys. {\bf #1}, #2 (#3)}

\def\pre#1#2#3{Phys. Rev. E {\bf #1}, #2 (#3)}

\def\pra#1#2#3{Phys. Rev. A {\bf #1}, #2 (#3)}
\def\physd#1#2#3{Physica D {\bf #1}, #2 (#3)}

\def\pr#1#2#3{Phys. Rev. {\bf #1}, #2 (#3)}

\def\physd#1#2#3{Physica D {\bf #1}, #2 (#3)}

\def\chaos#1#2#3{Chaos {\bf #1}, #2 (#3)}
\def\epl#1#2#3{Europhys. Lett. {\bf #1}, #2 (#3)}
\def\nld#1#2#3{Nonlinear Dyn. {\bf #1}, #2 (#3)}

\def\srep#1#2#3{Sci. Rep. {\bf #1}, #2 (#3)}
\def\pr#1#2#3{Phys. Rep. {\bf #1}, #2 (#3)}

\def\ep{\varepsilon}

\def\ie{i.e. }

\def\etal{{\it et al.}}

\def\beqr{\begin{eqnarray}}
\def\eqnr{\end{eqnarray}}
\def\beqrs{\begin{eqnarray*}}
\def\eqnrs{\end{eqnarray*}}
\def\beq{\begin{equation}}
\def\bc{\begin{center}}
\def\ec{\end{center}}
\def\eqn{\end{equation}\noindent}
\begin{document}
\title{Explosive synchronization in coupled stars}
\author{Ruby Varshney$^1$, Kaustubh Manchanda$^{2,3}$ and Haider Hasan Jafri$^1$}
\affiliation{$^1$Department of Physics, Aligarh Muslim University, Aligarh 202 002, India \\ $^2$School of Arts and Sciences, Azim Premji University, Bengaluru, Karnataka 562 125, India\\
$^3$College of Arts and Sciences, Abu Dhabi University, Abu Dhabi {59911}, UAE.}
\begin{abstract}
 We study the effect of network topology on the collective dynamics of an oscillator ensemble. Specifically, we explore explosive synchronization in a system of interacting star networks. Explosive synchronization is characterized by an abrupt transition from an incoherent state to a coherent state. 
 In this study, we couple multiple star networks through their hubs and study the emergent dynamics as a function of coupling strength. The dynamics of each node satisfies the equation of a Kuramoto oscillator. We observe that for a small inter-star coupling strength, the hysteresis width between the forward and backward transition point is minimal, which increases with an increase in the inter-star coupling strength. This observation is independent of the size of the network. Further, we find that the backward transition point is independent of the number of stars coupled together and the inter-star coupling strength, which is also verified using the Watanabe and Strogatz (WS) theory.
\end{abstract}
\maketitle

\begin{section}{Introduction}
Synchronization is an important phenomenon observed in the study of networked dynamical systems~\cite{Pikovsky-Book, Boccaletti-Book}. In the context of an ensemble of interacting units, synchronization refers to a transition from an incoherent state to a coherent state. Such an onset of coherence in real systems has been reported in power grids~\cite{dorfler2012,motter2013}, neuronal networks~\cite{cobb1995}, communication networks~\cite{ling}, and circadian rhythms~\cite{winfree1987}. Recently, special attention has been paid to understand the properties of complex networks consisting of dynamical units on the nodes~\cite{Watts-1998, Barabasi-1999, Boccaletti-2006}. These studies have revealed that the underlying topology plays a crucial role in the onset of synchronization~\cite{Boccaletti-2006, Moreno-2004, Zhou-2006, arenas}. Most studies have reported that the transition from desynchrony to synchrony is second-order in nature. This is characterized by an order parameter showing a smooth transition with changing coupling strength. 

With the work of Gardenes \etal~\cite{gardenes}, it was observed that in the case of a scale-free network, the order parameter might change discontinuously, resulting in the so-called explosive synchronization (ES). This transition is characterized by a well-defined hysteresis associated with the backward and forward transitions. It has been found to occur as a result of the correlation between the natural frequency of the dynamical units and the network topology in the case of a scale-free network. Many networks in nature are heterogeneous and have scale-free topology where a few nodes, namely the hubs, hold most connections. In such a scenario, a typical motif is a star network. There have been several studies to understand the collective phenomenon in the case of star networks, namely, the chimera state~\cite{meena, muni}, pattern formation~\cite{nikos,shena} and explosive synchronization~\cite{gardenes,coutinho,zou,pavel,vlasov,yusra2024}. In Ref.~\cite{gardenes,vlasov,yusra2024}, ES in a single star has been studied by introducing the degree-frequency correlation.

However, network synchronization can be influenced by the presence of other networks. In many realistic situations, it was observed that the network may display several levels of topological organization, which is not exactly described by typical network models~\cite{Song-2005}. In the case of the cerebral cortex of mammalian brains, the connectivity is organized as subnetworks of interacting neurons~\cite{Zhou-2006, Gardenes-2010}.  In transportation and telecommunication systems, the units that serve as transshipment and switching points may be regarded as a system of interacting hubs~\cite{Campbell-1994, Kim-2008}. Thus, it will be interesting to understand the collective dynamics in a situation wherein the underlying topology consists of a few central elements, namely the hubs that are interacting with each other. To simplify the situation, one can consider a star network that evolves under the influence of other star networks~\cite{xu, zhu2021}. Collective dynamics in the case of multilayer networks, where one layer might affect the dynamics of the other layer, has also been explored in Refs.~\cite{boccaletti2014,leyva,jalan23, zhang}.

In this study, we consider a network of two or more coupled star networks and study the phenomenon of explosive synchronization in the presence of degree-frequency correlation. 
We derive analytically a criterion for the synchronization of two identical star networks. This analysis is further extended to study the transitions in case of $N$ coupled star networks.
The analytical results are then compared with the numerical findings and we observe that the numerical and analytical results are in good agreement with each other. 
We note that the backward transition point is independent of the coupling strength and the size of the network; however, we predict an upper bound for the forward transition points. Throughout the investigation, we make use of the fourth-order Runge-Kutta method to solve the differential equations. 

This report is organized as follows: In Sec.~\ref{sec-2stars}, we study the dynamics in the case of two coupled star networks. We also give analytical support to the numerical findings for two coupled stars. In Sec.~\ref{sec:nstars}, we describe the collective dynamics in situations where we have three or more star networks interacting through their hubs. We give an analytical prescription to find the backward transition in the case of the $N$-coupled star networks. Finally, we summarize the findings in Sec.~\ref{sec:discuss}.

 \begin{figure}
   \includegraphics[width = 60mm,height=30mm]{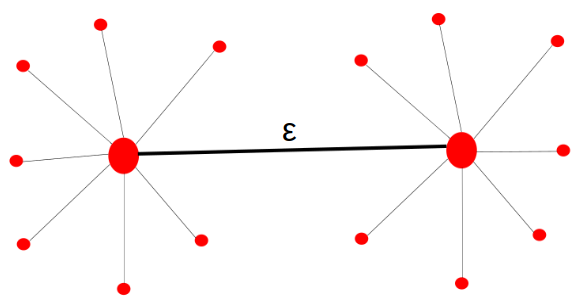}
   \caption{Two-star networks coupled via hubs with the inter-star coupling strength ($\ep$).}
    \label{fig0}
   \end{figure}
   
\end{section}

\begin{section}{Two coupled star networks}
\label{sec-2stars}
\subsection{Numerical Results}

 Consider a system of two coupled star networks (Fig.~\ref{1}) interacting through their hubs. The dynamics of each node is modelled via a Kuramoto oscillator. The equations of motion for the system are given by
\begin{equation}
	\label{1}
	\begin{aligned}
	\dot{\phi}_j &= \omega+\lambda_1\sin(\phi_h-\phi_j),  \qquad 1 \leq j \leq N_1   \\
     \frac{1}{\beta}\dot{\phi}_h &= \omega+\frac{\lambda_1}{N_1} \sum_{j=1}^{N_1} \sin(\phi_j-\phi_h)+\ep \sin(\phi'_h-\phi_h), \\
      \dot{\phi}'_j &= \omega+\lambda_2\sin(\phi'_h-\phi'_j), \qquad 1 \leq j \leq N_2 \\
     \frac{1}{\beta}\dot{\phi}'_h &= \omega+\frac{\lambda_2}{N_2} \sum_{j=1}^{N_2} \sin(\phi'_j-\phi'_h)+\ep \sin(\phi_h-\phi'_h).
	\end{aligned}
\end{equation}
where $\phi_j$ ($\phi'_j$) and $\phi_h$ ($\phi'_h$) are the phases of the oscillator at the nodes and at the hub, respectively, of the first (second) star. $\lambda_i$'s are the intra-star coupling strengths and $\ep$ denotes the inter-star coupling strength. All the oscillators have the same frequency \ie $\omega=1$, and the mismatch parameter is set to $\beta=10$. $N_1$ and $N_2$ represent the number of  nodes in each star. To understand the collective dynamics of the system, we have calculated the global order parameter $R$, given by
\beqr
\centering
\label{2}
R &=& \frac{1}{N} \left\lvert \sum_{j=1} ^{N} e^{i \phi_j}\right\rvert,
\eqnr
where $N$ ($=N_1+N_2+2$) is the total number of oscillators in the system.
We consider a situation where both the stars have identical intra-star coupling strengths, \ie $\lambda_1=\lambda_2=\lambda$ and study the variation of the global order parameter $R$ as a function of $\lambda$. In Fig.~\ref{fig:1}, we plot the global order parameter for different values of the inter-star coupling parameter ($\ep$). For any given value of $\ep$, we observe that the system shows a first-order transition to synchrony. As shown in Fig.~\ref{fig:1}, the order parameter changes discontinuously for both the forward and backward continuations as $\lambda$ varies. Since the backward and forward transition points are distinct, the system shows a well-defined hysteresis. The hysteresis width increases as the coupling strength $\ep$ varies. As the inter-star coupling is switched on, we observe a transition in the order parameter with a non-zero hysteresis width. We plot one such value in Fig.~\ref{fig:1} for $\ep=0.01$, for which the hysteresis width is very small. If $\ep$ is increased further ($\ep=0.1,0.3,0.5$), the hysteresis width increases as shown in Fig.~\ref{fig:1}. Note that the increase in hysteresis width occurs as a result of the shift in the forward transition point while the backward transition point is fixed for different values of $\ep$. 

\begin{figure}
 \includegraphics[width = 70mm,height=80mm, angle=270]{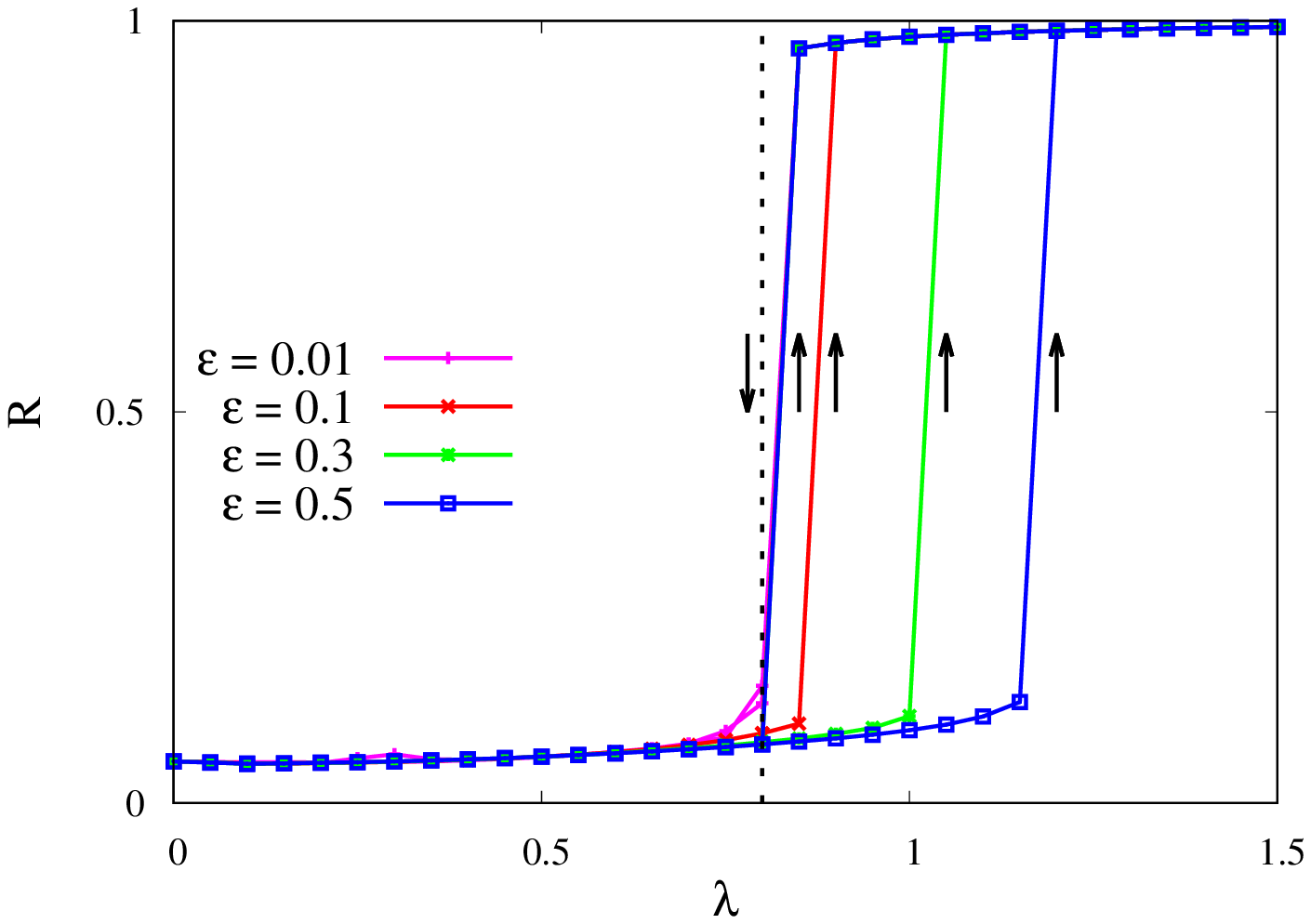}
\caption{Variation of global order parameter ($R$) with intra-star coupling strength $\lambda$ in forward and backward directions for two coupled star networks, with the number of oscillators in each star as 100 at different values of inter-star coupling strength $\ep$. The results are independent of the number of nodes in each star and have been numerically verified for $N_1=N_2 = 100$ and $500$.}
\label{fig:1}
\end{figure}

\subsection{Analytical Results}
\label{sec:ana2stars}
To understand the dynamics, we make use of the Watanabe-Strogatz (WS) approach~\cite{Watanabe-1993, Watanabe-1994} by introducing phase differences $\theta_j, \theta'_j$ and $\theta_h$ as follows
\begin{equation}
\label{3}
\begin{aligned}
    \theta_j = \phi_j - \phi_h, \quad
    \theta'_j = \phi'_j - \phi'_h, \quad
    \theta_h = \phi'_h - \phi_h.  
\end{aligned}
\end{equation}
In terms of new variables (Eq.~\eqref{3}) along with $N_1=N_2=N'$, the system Eqs.~\eqref{1} can be written as,
\begin{equation}
\label{4}
    \begin{aligned}
        \dot{\theta}_j &= \omega(1-\beta) -\lambda \sin{\theta_j} - \frac{\beta \lambda}{N'} \sum_{l=1}^{N'} \sin{\theta_l} - \beta \ep \sin{\theta_h},\\
          \dot{\theta}'_j &= \omega(1-\beta) -\lambda \sin{\theta'_j} - \frac{\beta \lambda}{N'} \sum_{l=1}^{N'} \sin{\theta'_l} + \beta \ep \sin{\theta_h},\\
          \dot{\theta}_h &= \frac{\beta \lambda}{N'} \left[\sum_{l=1}^{N'} \sin{\theta'_l} - \sum_{l=1}^{N'} \sin{\theta_l}\right] - 2\beta\ep \sin{\theta_h}
    \end{aligned}
\end{equation}
Note that the terms $\frac{1}{N'}\sum_{l=1}^{N'} \sin{\theta_l}$ and $\frac{1}{N'}\sum_{l=1}^{N'} \sin{\theta'_l}$ can be written as
\begin{equation}
\label{5}
    \begin{aligned}
        \frac{1}{N'}\sum_{l=1}^{N'} \sin{\theta_l} = \operatorname{Im}\left[  \frac{1}{N'}\sum_{l=1}^{N'} e^{\di \theta_l}\right]  = \operatorname{Im}(R_1) ,\\
        \frac{1}{N'}\sum_{l=1}^{N'} \sin{\theta'_l} = \operatorname{Im}\left[  \frac{1}{N'}\sum_{l=1}^{N'} e^{\di \theta'_l}\right]  = \operatorname{Im}(R_2).
    \end{aligned}
\end{equation}
where $R_1$ and $R_2$ are the local order parameters. Substituting Eq.~\eqref{5} in Eq.~\eqref{4} with $\Delta \omega= \omega(1-\beta)$, we get
\begin{equation}
\label{6}
    \begin{aligned}
        \dot{\theta}_j &= \Delta \omega+\lambda \operatorname{Im}(e^{-\di \theta_j}) - \beta \lambda \operatorname{Im}(R_1) - \beta \ep \sin{\theta_h},\\
          \dot{\theta}'_j &= \Delta \omega+\lambda \operatorname{Im}(e^{-\di \theta'_j}) - \beta \lambda \operatorname{Im}(R_2) + \beta \ep \sin{\theta_h},\\
          \dot{\theta}_h &= \beta \lambda \left[\operatorname{Im}(R_2)-\operatorname{Im}(R_1)\right]- 2\beta\ep \sin{\theta_h}
    \end{aligned}
\end{equation}
This higher dimensional equation can be reduced to a lower dimensional equation using WS ansatz \cite{Watanabe-1993, Watanabe-1994} for the general form
\begin{equation}
\label{7}
\dot{\theta}_k = g(t) + \operatorname{Im}(G(t) e^{-\di \theta_k}).
\end{equation}
 By comparing Eq.~\eqref{6} with Eq.~\eqref{7}, we get
 \begin{equation}
 \label{8}
  \begin{aligned}
  g(t) &= \Delta \omega-\beta \ep \sin{\theta_h}- \beta\lambda \operatorname{Im}(R_1); \quad G(t) = \lambda, \\
  g'(t) &= \Delta \omega+\beta \ep \sin{\theta_h}- \beta\lambda \operatorname{Im}(R_2); \quad G'(t) = \lambda
  \end{aligned}   
   \end{equation}
The WS transformation from the phase variables $\theta_k$ to a set of global variables $z,\alpha$ ($z$ being a complex and $\alpha$ being a real variable) changes Eq.~\eqref{7} to
   \begin{equation}
   \label{10}
       \begin{aligned}
           \dot{z} &= \di g(t) z + \frac{G(t)}{2} - \frac{G(t)^*}{2}z^2, \\
           \dot{\alpha}&=g(t)+\operatorname{Im}\left[z^* G(t)\right]
        \end{aligned}
   \end{equation}

With WS transformation applied for the two-star network, we are left with variables $z,\alpha,z',\alpha'$ and $\theta_h$ whose variation is given as
\begin{equation}
\label{11}
    \begin{aligned}
        \dot{z}&=\di \left[\Delta\omega-\beta \ep \sin{\theta_h}- \beta\lambda \operatorname{Im}(z)\right]z + \frac{\lambda}{2}\left(1-z^2\right),\\
        \dot{\alpha}&= \Delta\omega-\beta \ep \sin{\theta_h}- \beta\lambda \operatorname{Im}(z)+\lambda\operatorname{Im}(z^*),\\
        \dot{z'}&=\di \left[\Delta\omega+\beta \ep \sin{\theta_h}- \beta\lambda \operatorname{Im}(z')\right]z' + \frac{\lambda}{2}\left(1-{z'}^2\right),\\
        \dot{\alpha}'&= \Delta\omega+\beta \ep \sin{\theta_h}- \beta\lambda \operatorname{Im}(z')+\lambda\operatorname{Im}({z'}^*),\\
         \dot{\theta}_h &= \beta \lambda \left[\operatorname{Im}(z')-\operatorname{Im}(z)\right]- 2\beta\ep \sin{\theta_h}
    \end{aligned}
\end{equation}
 where $z,z'$ are the modified order parameters corresponding to $R_1,R_2$ respectively.
Since the equations for $z,z'$ do not contain any $\alpha, \alpha'$ terms, we can solve them independently along with the equation for $\theta_h$ as follows
\begin{equation}
\label{12}
    \begin{aligned}
        \dot{z}&=\di \left[\Delta\omega-\beta \ep \sin{\theta_h}- \beta\lambda \operatorname{Im}(z)\right]z + \frac{\lambda}{2}\left(1-z^2\right),\\
        \dot{z'}&=\di \left[\Delta\omega+\beta \ep \sin{\theta_h}- \beta\lambda \operatorname{Im}(z')\right]z' + \frac{\lambda}{2}\left(1-{z'}^2\right),\\
         \dot{\theta}_h &= \beta \lambda \left[\operatorname{Im}(z')-\operatorname{Im}(z)\right]- 2\beta\ep \sin{\theta_h}
    \end{aligned}
\end{equation}
The complex variables $z$ and $z'$ can be written as,
\begin{equation}
\label{13}
    \begin{aligned}
        z = \frac{1}{N'}\sum_{l=1}^{N'} e^{\di \theta_l} = \rho_1 e^{\di \psi_1} ,\\
        z' = \frac{1}{N'}\sum_{l=1}^{N'} e^{\di \theta'_l} = \rho_2 e^{\di \psi_2} 
    \end{aligned}
\end{equation}

Substituting Eq.~\eqref{13} in Eq.~\eqref{12}, we get
\begin{equation}
\label{14}
    \begin{aligned}
        \dot{\rho}_1 &= \frac{\lambda}{2} \left(1-\rho_1^2\right) \cos{\psi_1},\\
        \dot{\psi}_1 &= \Delta\omega- \beta \ep \sin{\theta_h} - \beta \lambda \rho_1 \sin{\psi_1}-\frac{\lambda}{2 \rho_1}\left(1+\rho_1^2\right) \sin{\psi_1} ,\\
          \dot{\rho_2} &= \frac{\lambda}{2} \left(1-\rho_2^2\right) \cos{\psi_2},\\
        \dot{\psi}_2 &= \Delta\omega+ \beta \ep \sin{\theta_h} - \beta \lambda \rho_2 \sin{\psi_2}-\frac{\lambda}{2 \rho_2}\left(1+\rho_2^2\right) \sin{\psi_2},\\
         \dot{\theta}_h &= \beta \lambda \left[\rho_2\sin{\psi_2}-\rho_1\sin{\psi_1}\right]- 2\beta \ep \sin{\theta_h}
    \end{aligned}
\end{equation}
For steady state,
\begin{equation}
\label{15}
    \begin{aligned}
        \dot{\rho}_1 &=\dot{\rho}_2=0,\\
\implies \rho_1&=\rho_2=1 \quad \qquad \text{(synchronous branch),}\\
\text{or} \quad \cos{\psi_1}&=\cos{\psi_2}= 0\quad \text{(asynchronous branch).}
    \end{aligned}
        \end{equation}
We shall now consider the cases to find the solutions for synchronous and asynchronous branches separately.
\subsubsection{Synchronous branch}
The stability of the synchronous branch gives the backward transition point of the explosive synchronization, so for $\rho_1=\rho_2=1$, Eq.~\eqref{14} becomes,

\begin{equation}
\label{16}
    \begin{aligned}
        \dot{\psi}_1 &= \Delta\omega- \beta \ep \sin{\theta_h} - \beta \lambda \sin{\psi_1}-\lambda\sin{\psi_1} ,\\
        \dot{\psi}_2 &= \Delta\omega+ \beta \ep \sin{\theta_h} - \beta \lambda\sin{\psi_2}-\lambda\sin{\psi_2},\\
         \dot{\theta}_h &= \beta \lambda \left[\sin{\psi_2}-\sin{\psi_1}\right]- 2\beta \ep \sin{\theta_h}
    \end{aligned}
\end{equation}

Since for the synchronized state, all the hubs are in the same state (\ie $\theta_h=0$) and putting $\dot{\psi}_1=0$, we obtain
\begin{equation}
    \label{17}
    \begin{aligned}
        \sin{\psi_1}=\frac{\Delta \omega}{\lambda (\beta+1)},\\
        \left|\frac{-\omega(\beta-1)}{\lambda(\beta+1)}\right| \leq 1.
    \end{aligned}
\end{equation}
Thus, the synchronous branch is stable for
\begin{equation}
\label{17a}
\begin{aligned}
    \lambda \geq \frac{\omega(\beta-1)}{(\beta+1)}. \\
     \implies \lambda_c^b = \frac{\omega(\beta-1)}{(\beta+1)}
    \end{aligned}
 \end{equation}
 
From Eq.~\eqref{17a}, we note that the backward transition point is independent of the inter-star coupling parameter ($\ep$). For $\omega=1$ and $\beta=10$, this transition point is found to occur at $\lambda^b_c=0.8$, as shown by the black dashed line in Fig.~\ref{fig:1}. This is found to be in good agreement with the numerical results (Fig.~\ref{fig:1}).

\subsubsection{Asynchronous branch}
The stability of the asynchronous branch gives the forward point of the explosive transition, so we put $\psi_1=\psi_2= \frac{\pi}{2}$ in Eq.~\eqref{14}, 
\begin{equation}
    \label{18}
    \begin{aligned}
         \dot{\psi}_1 &= \Delta\omega- \beta \ep \sin{\theta_h} - \beta \lambda \rho_1-\frac{\lambda}{2 \rho_1} \left(1+\rho_1^2\right) ,\\
        \dot{\psi}_2 &= \Delta\omega+ \beta \ep \sin{\theta_h} - \beta \lambda \rho_2-\frac{\lambda}{2 \rho_2} \left(1+\rho_2^2\right) ,\\
         \dot{\theta}_h &= \beta \lambda \left[\rho_2 - \rho_1\right]- 2\beta\ep \sin{\theta_h}
    \end{aligned}
\end{equation}
Put $\dot{\psi}_1=\dot{\psi}_2=0$ and calculate the value of $\rho_1$ from Eq.~\eqref{18} as,

\begin{equation}
    \label{19}
    \rho_1^2 (2\beta+1)\lambda+2\left[\omega(\beta-1)+\beta \ep \sin{\theta_h}\right]\rho_1 + \lambda =0
\end{equation}
Now, $\rho_1$ has to be real, so the determinant should be positive.
\beqr
    \label{20}
      \left[\omega(\beta-1)+\beta \ep \sin{\theta_h}\right]^2 - \lambda^2 (2\beta+1) \geq 0, \nonumber \\
    \sin{\theta_h} \geq \frac{\lambda \sqrt{2\beta+1}-\omega(\beta-1)}{\beta \ep}, \nonumber \\
     1 \geq \sin{\theta_h} \geq \frac{\lambda \sqrt{2\beta+1}-\omega(\beta-1)}{\beta \ep}, \nonumber \\  
     1 \geq \frac{\lambda \sqrt{2\beta+1}-\omega(\beta-1)}{\beta \ep}, \nonumber \\
     \lambda \leq \frac{\beta\ep+\omega(\beta-1)}{\sqrt{2\beta+1}} \nonumber \\
  \implies  \lambda_c^f = \frac{\beta\ep \pm \omega(\beta-1)}{\sqrt{2\beta+1}}   
\eqnr

This gives the upper bound for the forward transition point. The forward transition point depends on the coupling parameter $\ep$. Numerically, the forward transition point for each $\ep$ is found to be below $\lambda_c^f$ given by  Eq.~\eqref{20} (c.f. Fig.~\ref{fig:1}).

 \end{section}

\begin{section}{N-coupled Star Network (N $>$ 2)}
\label{sec:nstars}
 \begin{figure}[htp]
{\includegraphics[width = 60mm,height=50mm]{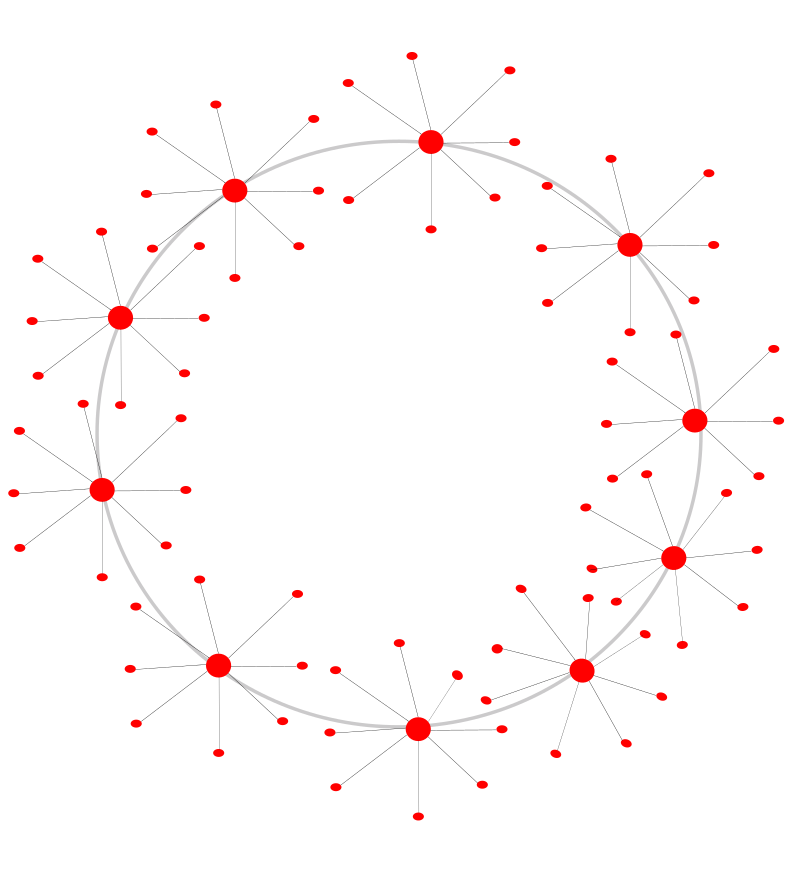}} 
\caption{ Closed arrangement of ten-star network coupled via hub oscillator with two hub-to-hub connections.}
\label{fig:4a}
\end{figure}

 In this section, we extend our analysis to a more general scenario. We consider a system of $N$ coupled star networks. The connection topology is that of a ring, where we place the hub of a given star network on each node of a ring. The ring has $N$ such nodes. Thus, the hub of each star network is connected to the hubs of the two other star networks. The schematic diagram in Fig.~\ref{fig:4a} describes a system of $10$ star networks interacting on a ring topology. The equations of motion of $N$ such coupled star networks are given by

\beqr
\label{41a}
	  \dot{\phi}_{j}^k & =& \omega+\lambda_j\sin(\phi_h^k-\phi_{j}^k), \hspace{1cm}  1\leq j \leq N_j \nonumber\\
     \frac{1}{\beta}\dot{\phi}_h^k &=& \omega+\frac{\lambda_j}{N_j} \sum_{j=1}^{N_j} \sin(\phi_{j}^k-\phi_{h}^k)\nonumber\\
&& +\frac{\ep}{2}\sum_{i=1}^{N}\left\lbrace A(i,k) \sin(\phi_{h}^i-\phi_{h}^k)\right\rbrace
\eqnr
where the superscripts $k, i = 1, 2 \dots N~ (k \ne i)$ represent each star network and the subscript $j=1, 2 \dots N_j$ represents the nodes of an individual star network. Thus, ${\phi_{j}^k}$ is the phase of the $j^{th}$ oscillator in the $k^{th}$ star network and ${\phi_{h}^k}$ is the phase of the hub in the $k^{th}$ star network. $A(i,k)$ is the adjacency matrix that describes the coupling between the hubs of the star networks $i$ and $k$. $A(i,k)=1$ when there is a connection between the hub of the $i^{th}$ star and the hub of the $k^{th}$ star, otherwise it is equal to zero.
Now, one can generalize the analysis done in Sec.~\ref{sec:ana2stars}, by introducing the phase differences given by
     \beqr
        \label{defnewvar-k}
        \theta_j^k &=& \phi_j^k - \phi_h^k, \nonumber \\ 
        \theta_h^k &=& \phi_h^{k+1} - \phi_h^k, 
    \eqnr
   For the star network $k$, the order parameter $R_k$ may be defined as (c.f. Eq.~\eqref{5}) 

\begin{equation}
\label{rk}
    \begin{aligned}
        \frac{1}{N_k}\sum_{l=1}^{N_k} \sin{\theta_l^k} = \operatorname{Im}\left[  \frac{1}{N_k}\sum_{l=1}^{N_k} e^{\di \theta_l^k}\right]  = \operatorname{Im}(R_k) 
    \end{aligned}
\end{equation}
 Rewriting Eq.~\eqref{41a} in terms of new phase variables (Eq.~\eqref{defnewvar-k}) and the order parameter $R_k$ (Eq.~\eqref{rk}), we obtain
\beqr
\label{42}
    \dot{\theta}^k_j &=& \Delta \omega + \lambda \operatorname{Im}(e^{-\di \theta_j^k}) -\beta \lambda \operatorname{Im}(R_k) \nonumber \\ && - \frac{\beta \ep}{2} \left[\sin{\theta_h^k}-\sin{\theta_h^{k-1}} \right],\nonumber \\
    \dot{\theta}^k_h &=& \beta \lambda \left[\operatorname{Im}(R_{k+1})-\operatorname{Im}(R_k) \right] \nonumber \\ && + \frac{\beta \ep}{2} \left[\sin{\theta_h^{k+1}}+\sin{\theta_h^{k-1}} - 2\sin{\theta_h^k}\right], \nonumber \\ 
   \eqnr

   Comparing Eq.~\eqref{42} with the standard equation
   \beq
     \dot{\theta_j^k} = g_k(t) + \operatorname{Im} \left[G_k(t) e^{-\di \theta_j^k}\right],
   \eqn
   we get,
   \beqr
     g_k &=& \Delta \omega  -\beta \lambda \operatorname{Im}(R_k)  - \frac{\beta \ep}{2} \left[\sin{\theta_h^k}-\sin{\theta_h^{k-1}} \right], \nonumber \\ 
     G_k &=& \lambda \nonumber \\ 
   \eqnr

The variation of the complex global variables ($z_k$) can be written as
\beqr
    \dot{z_k} = \di g_k G_k + \frac{G_k}{2} - \frac{G_k^{*} z_k^2}{2} ; \quad k = 1,2...N
\eqnr
  Put 
  \beq
  z_k = \rho_k e^{\di \psi_k}
  \eqn
  In terms of $\rho_k$ and $\psi_k$, Eqs.~\eqref{42} are given by
  \beqr
  \label{43}
    \dot{\rho}_k &=& \frac{\lambda}{2} \left(1-\rho_k^2    \right) \cos{\psi_k}, \nonumber \\
    \dot{\psi}_k &=& \Delta\omega-\frac{\beta \ep}{2} \left[\sin{\theta_h^{k+1}}-\sin{\theta_h^{k}} \right] - \beta \lambda \rho_k \sin{\psi_k} \nonumber \\ && -\frac{\lambda}{2 \rho_k}\left(1+\rho_k^2\right) \sin{\psi_k}, \nonumber \\
    \dot{\theta}^k_h &=& \beta \lambda \left[\rho_{k+1} \sin{\psi_{k+1}}-\rho_k \sin{\psi_k}\right] \nonumber \\ && + \frac{\beta \ep}{2} \left[\sin{\theta_h^{k+1}}+\sin{\theta_h^{k-1}} - 2\sin{\theta_h^k}\right]
  \eqnr
  For steady state,
  \beqr
  \label{sts}
    \dot{\rho}_k = \dot{\psi}_k=0
   \eqnr
From Eq.~\eqref{43}, the solution $\rho_k=1$ corresponding to $\dot{\rho}_k =0$ accounts for the stability of the synchronized state, and hence gives the backward transition point.  Put $\rho_k=1$ in Eq.~\eqref{43} and solve for $\psi_k$ as follows
 \beqr
 \label{44}
\Delta \omega -\frac{\beta \ep}{2} \left[\sin{\theta_h^{k+1}}-\sin{\theta_h^{k}} \right] - \beta \lambda  \sin{\psi_k}-\lambda \sin{\psi_k} = 0 \nonumber \\
 \eqnr
 In the synchronized state, all the hubs are in the same state, so the phase difference between any two hubs: 
 \beq
 \label{45}
 \theta_h^{k+1}=\theta_h^k=0
 \eqn
 
 Substituting Eq.~\eqref{45} in Eq.~\eqref{44}, 
 \beqr
   \sin{\psi_k} = \frac{-\omega (\beta-1)}{\lambda(\beta+1)}, \nonumber \\
   \left\lvert \frac{-\omega (\beta-1)}{\lambda(\beta+1)} \right\rvert \leq 1, \nonumber \\
   \lambda_c^b \geq \frac{\omega (\beta-1)}{(\beta+1)} .
   \label{eqln}
   \eqnr

The minimum coupling strength for which the synchronized branch remains stable can be obtained by using the analysis done for a two-coupled star network. The synchronous branch is stable after $\lambda=0.8$ and is independent of the inter-star coupling~($\epsilon$) strength and the number of stars~($N$) coupled together.

For the asynchronous branch, we again put $\dot{\psi_k}=0$ (Eq.~\eqref{sts}) with $\psi_k=\pi/2$ in Eq.~\eqref{43}, we get

\beq
 \rho_k^2(2\beta+1)\lambda+\left[2 \omega(\beta-1)+\beta \ep\left(\sin{\theta_h^{k+1}}-\sin{\theta_h^k}\right)\right]\rho_k+\lambda =0
 \eqn
The determinant of the roots of the above quadratic equation should be positive, and it will lead to
 \beqr
 \sin{\theta_h^{k+1}}-\sin{\theta_h^k} \geq \frac{2 \lambda \sqrt{2\beta+1}-2 \omega (\beta-1)}{\beta \epsilon}
 \eqnr
The difference of the sine terms can be a maximum of $2$. This way, the stability of the asynchronous branch is achieved in the case of the $N$-coupled star networks. This gives an upper bound for the forward transition points.

\begin{figure}[b]
 \includegraphics[width = 70mm,height=80mm, angle=270]{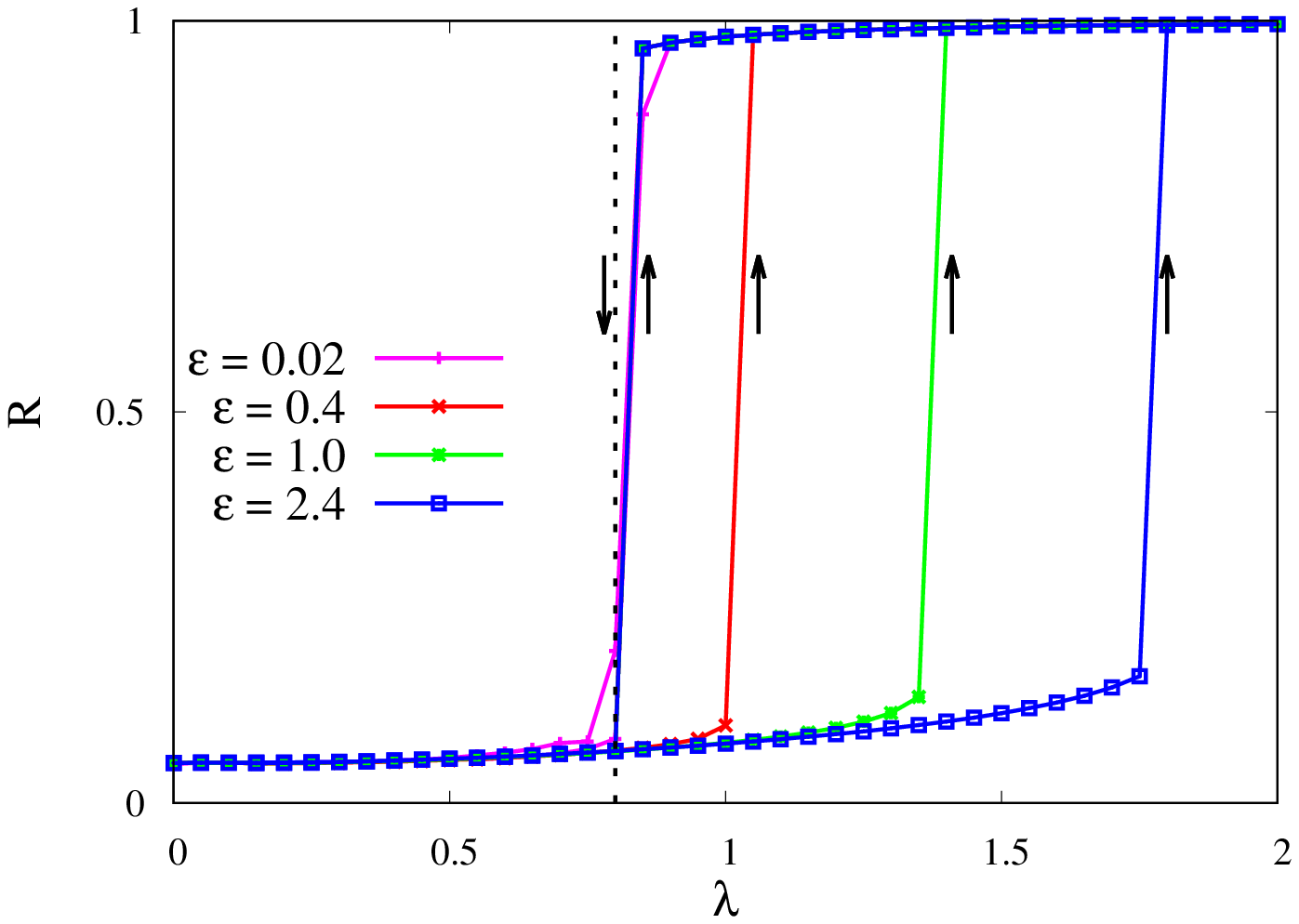}
\caption{ Variation of the global order parameter $R$ as a function of intra-star coupling for different values of the inter-star couplings ($\ep$) for $N=3$ (Eq.~\eqref{41a}). The parameter values are taken to be $\omega=1, \beta=10$ and $N_j=100$ number of leaves in each star at $\ep = 0.02$ (magenta), $\ep= 0.4$ (red), $\ep=1$ (green), and $\ep= 2.4$ (blue). The theoretically predicted backward transition point has been shown by the black dashed line.}
\label{fig:3}
\end{figure}

Now we shall numerically explore a situation when there are more star networks coupled together $(N>2)$ and study the order parameter as a function of the intra-star coupling strength for a fixed inter-star coupling. We consider a network of three coupled stars $(N=3)$, for which the equations of motion are given by Eq.~\eqref{41a} by setting $N=3$. The variation of the global order parameter $R$ as a function of $\lambda$ is shown in Fig.~\ref{fig:3}. We observe that the order parameter changes abruptly as the intra-star coupling ($\lambda$) is varied in the forward and the backward directions. Since the forward and backward transition points are distinct, the system exhibits a well-defined hysteresis. We also note that as the value of $\ep$ increases, the hysteresis width increases. At $\ep=0.02$, the hysteresis width is very small, which increases as the value of $\ep$ is increased, as shown by the red line ($\ep=0.4$), green line ($\ep=1$) and the blue line ($\ep=2.4$). Moreover, the backward transition point is exactly the same as predicted by Eq.~\eqref{eqln}. This is shown by a dashed line in Fig.~\ref{fig:3}. 

Further, we repeated the analysis for a bigger network taking $N=10$ (Eq.~\eqref{41a}) and found that the results are qualitatively the same as shown in Fig.~\ref{fig:4}. Again, we note that the forward and the backward transition points are distinct, resulting in well-defined hysteresis. We show this by plotting the order parameter for  $\ep=0.02, 0.3, 0.4, 0.5$ as shown in Fig.~\ref{fig:4}. The backward transition is the same and matches well with the analytically predicted point given by Eq.~\eqref{eqln}.
\end{section}

\begin{figure}[t]
 \includegraphics[width = 70mm,height=80mm, angle=270]{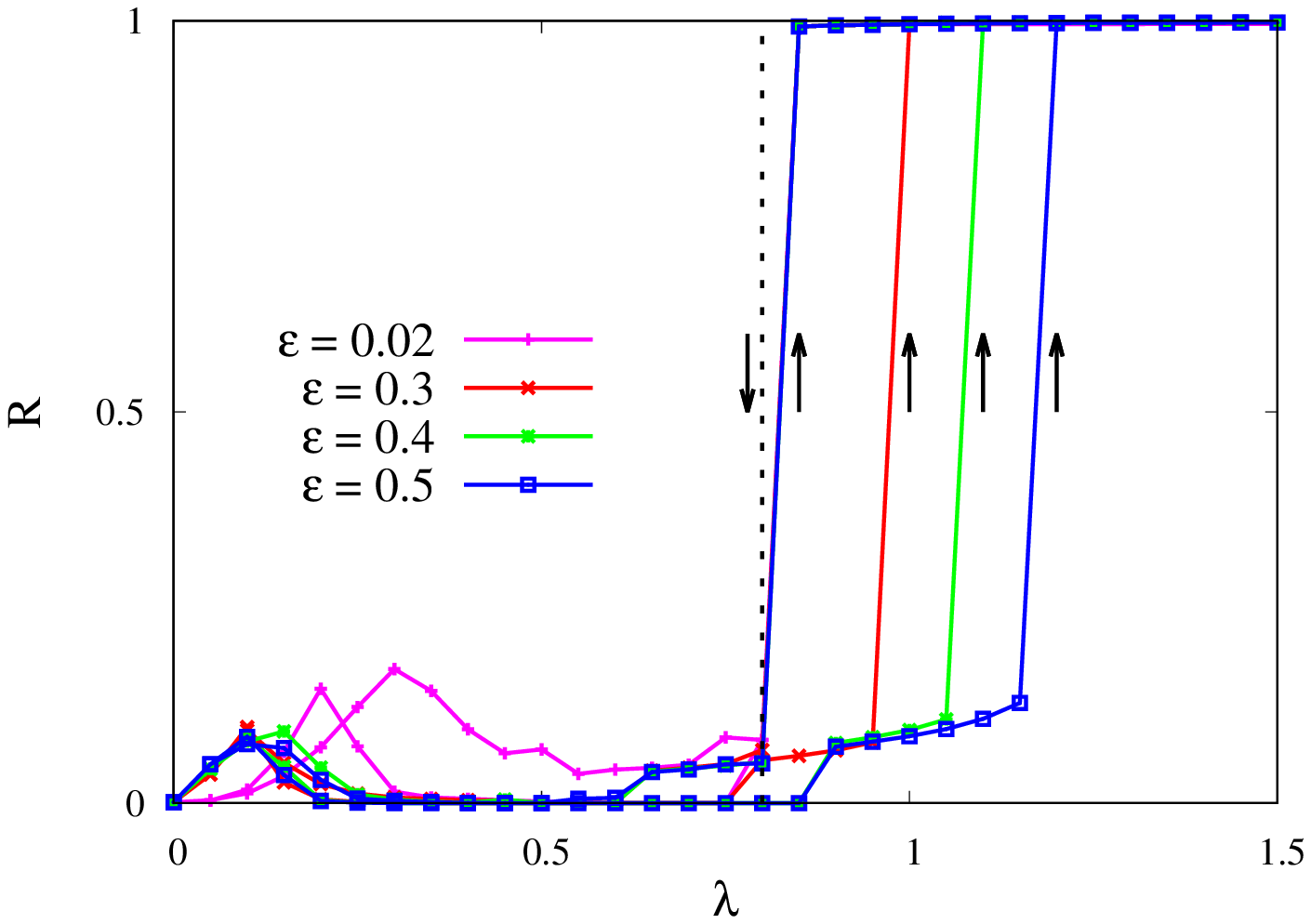}
\caption{Variation of the global order parameter ($R$) in case of ten coupled stars (N=10) given by Eq.~\eqref{41a} with number of leaves in each star $N_1=N_2=N_3 .... = N_{10} = 100$ at different $\ep$-values.}
\label{fig:4}
\end{figure}

\begin{section} {Discussion}
\label{sec:discuss}
     In this work, we study the collective dynamics of multiple star networks that are coupled through their hubs. We observe that the system exhibits a global first-order transition as the intra-star coupling ($\lambda$) increases in the forward and backward directions for different values of the inter-star coupling ($\ep$). As we increase $\ep$, the hysteresis width increases. The forward transition point (FTP) shifts towards the higher $\lambda$ value while the backward transition point (BTP) remains the same, thereby changing the hysteresis width. The BTP is found to be independent of the number of stars coupled together and the strength of the inter-star coupling. We have shown results for two, three and ten-coupled stars. These findings have been verified through the analytical treatment of the system. The dependency of the forward transition point on the inter-star coupling strength is obtained by the WS analysis. We analytically calculate the upper bound of the forward transition point and note that the numerically observed transition points are below the predicted value. In the case of BTP, the analytically predicted value is found to be in good agreement with the numerical findings.

 Our work highlights the role of topological connectivity in achieving synchronization in networks of heterogeneous degree distributions, namely star networks. Such network configurations have been known to form the backbone of several real world complex systems \cite{Bonifazi-2009, Bertolero-2018}. Unravelling the relationship between network architecture and emergent collective dynamics has been the focus of several studies found in the literature \cite{Singh-2011}. This work, in particular, elucidates the role of coupling via hubs in observing global synchrony. Similar studies have been performed on networks of chemical oscillators, where the authors have investigated the modes and mechanisms of achieving synchronization in systems coupled via hubs \cite{Sawicki-2021}. A system with interacting hubs, namely the central elements, appears in various fields, namely the communication systems, social networks and mammalian brain \cite{Zhou-2006, Gardenes-2010, Campbell-1994, Kim-2008}. The importance of hubs in relaying information and optimising network operation costs has also been discussed in several works \cite{Aykin-1995,Sawicki-2021,Campbell-1994,Kim-2008}. We, therefore, believe our results  have a many potential practical applications which could be a subject of future investigations.
     
Further, in the current study, we consider the Kuramoto model, which is analytically tractable and allows for fast numerical simulations. However, in more realistic situations, it will be interesting to consider more complicated dynamics to explore other dynamical features such as cluster synchronization, chimera states, relay synchronization etc.

\end{section}

\section*{ACKNOWLEDGMENTS}
RV wants to acknowledge CSIR, India, for the SRF fellowship under file no. 09/112(0601)/2018-EMR-I.

\end{document}